\documentclass[sigconf]{acmart}

\usepackage{balance}
\usepackage{longtable}
\usepackage{subcaption}
\usepackage{xcolor}
\usepackage{xspace}
\usepackage{soul}
\usepackage{multirow, graphicx}
\usepackage{listings}
\usepackage{amsmath}
\usepackage{appendix}
\usepackage{tabularx}
\usepackage{enumitem}
\usepackage{listings}
\usepackage{pdfcomment}
\usepackage{colortbl}
\copyrightyear{2024}
\acmYear{2024}
\setcopyright{acmlicensed}\acmConference[MSR '24]{21st International Conference on Mining Software Repositories}{April 15--16, 2024}{Lisbon, Portugal}
\acmBooktitle{21st International Conference on Mining Software Repositories (MSR '24), April 15--16, 2024, Lisbon, Portugal}
\acmDOI{10.1145/3643991.3644926}
\acmISBN{979-8-4007-0587-8/24/04}

\newcommand\ai{GPT-4\xspace}
\newcommand\aig{GPT-4 generated\xspace}

\newcommand\RqOne{RQ1}
\newcommand\RqTwo{RQ2}
\newcommand\RqThree{RQ3}
\newcommand\RqOneText{How well can code-stylometry features distinguish human-authored code from \aig code?}
\newcommand\RqTwoText{How influential are non-gameable features in differentiating human-authored vs. \aig code?}
\newcommand\RqThreeText{How well does the classifier perform when trained and evaluated on only correct solutions?}
\newcommand\RqFour{RQ4}
\newcommand\RqFourText{How well does the classifier perform when trained and evaluated across varying levels of problem difficulty?}

\newcommand{\numFeatures}{140\xspace}
\newcommand{\numProb}{399\xspace}
\newcommand{\numSol}{798\xspace}
\newcommand{\totalNumSol}{1596\xspace}
\newcommand{\numProbCorrect}{161 }

\definecolor{codegreen}{rgb}{0,0.6,0}
\definecolor{codegray}{rgb}{0.5,0.5,0.5}
\definecolor{codepurple}{rgb}{0.58,0,0.82}
\definecolor{backcolour}{rgb}{0.92,0.93,0.94}
\lstdefinestyle{codeliststyle}{
  backgroundcolor=\color{backcolour}, commentstyle=\color{codegreen},
  keywordstyle=\color{magenta},
  numberstyle=\tiny\color{codegray},
  stringstyle=\color{codepurple},
  basicstyle=\ttfamily\footnotesize,
  breakatwhitespace=false,         
  breaklines=true,                 
  captionpos=b,                    
  keepspaces=true,                 
  numbers=left,                    
  numbersep=5pt,    
  showlines=true
  showspaces=false,                
  showstringspaces=false,
  showtabs=false,                  
  tabsize=2
}
\begin{document}

\title{Whodunit: Classifying Code as Human Authored or \aig - A case study on CodeChef problems}
% \title{}

%%
%% The "author" command and its associated commands are used to define
%% the authors and their affiliations.
%% Of note is the shared affiliation of the first two authors, and the
%% "authornote" and "authornotemark" commands
%% used to denote shared contribution to the research.
\author{Oseremen Joy Idialu}
\email{j2idialu@uwaterloo.ca}
\orcid{0000-0003-3037-4241}
\affiliation{%
  \institution{University of Waterloo}
  \state{Ontario}
  \country{Canada}
}

\author{Noble Saji Mathews}
\email{noblesaji.mathews@uwaterloo.ca}
\orcid{0000-0003-2266-8848}
\affiliation{%
  \institution{University of Waterloo}
  \state{Ontario}
  \country{Canada}
}

\author{Rungroj Maipradit}
\email{rungroj.maipradit@uwaterloo.ca}
\orcid{0000-0003-4286-9807}
\affiliation{%
  \institution{University of Waterloo}
  \state{Ontario}
  \country{Canada}
}

\author{Joanne M. Atlee}
\email{jmatlee@uwaterloo.ca}
\orcid{0000-0002-0760-526X}
\affiliation{%
  \institution{University of Waterloo}
  \state{Ontario}
  \country{Canada}
}

\author{Meiyappan Nagappan}
\email{mei.nagappan@uwaterloo.ca}
\orcid{0000-0003-4533-4728}
\affiliation{%
  \institution{University of Waterloo}
  \state{Ontario}
  \country{Canada}
}

%%
%% By default, the full list of authors will be used in the page
%% headers. Often, this list is too long, and will overlap
%% other information printed in the page headers. This command allows
%% the author to define a more concise list
%% of authors' names for this purpose.
\renewcommand{\shortauthors}{Idialu et al.}

%%
%% The abstract is a short summary of the work to be presented in the
%% article.
\begin{abstract}
Artificial intelligence (AI) assistants such as GitHub Copilot and ChatGPT, built on large language models like \ai, are revolutionizing how programming tasks are performed, raising questions about whether code is authored by generative AI models. Such questions are of particular interest to educators, who worry that these tools enable a new form of academic dishonesty, in which students submit AI-generated code as their work. Our research explores the viability of using code stylometry and machine learning to distinguish between \aig and human-authored code. Our dataset comprises human-authored solutions from CodeChef and AI-authored solutions generated by \ai. Our classifier outperforms baselines, with an F1-score and AUC-ROC score of 0.91. A variant of our classifier that excludes gameable features (e.g., empty lines, whitespace) still performs well with an F1-score and AUC-ROC score of 0.89. We also evaluated our classifier on the difficulty of the programming problem and found that there was almost no difference
between easier and intermediate problems, and the classifier performed only slightly worse on harder problems. Our study shows that code stylometry is a promising approach for distinguishing between \aig code and human-authored code.
\end{abstract}

%%
%% The code below is generated by the tool at http://dl.acm.org/ccs.cfm.
%% Please copy and paste the code instead of the example below.
%%

%%
%% Keywords. The author(s) should pick words that accurately describe
%% the work being presented. Separate the keywords with commas.
%% A "teaser" image appears between the author and affiliation
%% information and the body of the document, and typically spans the
%% page.

%\received{20 February 2007}
%\received[revised]{12 March 2009}
%\received[accepted]{5 June 2009}

%%
%% This command processes the author and affiliation and title
%% information and builds the first part of the formatted document.
\maketitle

\section{Introduction}
AI tools like Github Copilot~\cite{copilot22}, ChatGPT~\cite{openai22}, and Code Whisperer~\cite{aws23} are disrupting how educators teach and assess programming. These tools are promoted as coding “assistants” that aim to improve developer productivity by suggesting code snippets, bug fixes, code refactorings, and test cases. The use of AI assistants in introductory programming courses has been shown to increase the productivity of novice programmers in solving introductory programming problems, both in terms of improving the quality of their programs and easing the cognitive load and development effort required~\cite{Kazemitabaar_CHI23}. 

Eventually, however, educators need to assess how well students can program \textit{without} the aid of their coding assistants, and they are worried about academic dishonesty. Programming courses already suffer from high levels of plagiarism~\cite{Albluwi2019} and contract cheating~\cite{ITiCSE2013}. The ease with which these new tools can automatically generate code raises concerns about a new form of academic dishonesty, in which students submit AI-generated programs as their work~\cite{ansley_ai_implications_2022, becker_ai_in_education_2023, Lau2023}. Existing approaches to detecting plagiarism among student-submitted programs rely on automated similarity comparison tools, but these tools are unlikely to detect AI-generated solutions because AI-generated code has low similarity to student-authored code ~\cite{puryear_copilot_in_class_2022}.

The goal of our research is to determine the viability of constructing a classifier that can distinguish between \aig and human-authored code. We hypothesize that the low similarity between student-authored and \aig code suggests that code stylometry and machine-learning classification can be used to distinguish between the two.

In this paper we make the following contributions: (1) We make a best-effort attempt at constructing a classifier for detecting \aig Python code, using a combination of supervised machine learning (XGBoost~\cite{xgboost_2016}), and a collection of \numFeatures code-stylometry features. The classifier is trained and evaluated on a dataset comprising \numSol human-authored solutions and \numSol \aig solutions to \numProb Python problems of varying degrees of difficulty from CodeChef.\footnote{\url{https://www.codechef.com/}} To our knowledge, this is the first attempt to construct a classifier for detecting \aig code based on a dataset with over 1000 solutions. (2) Our evaluation focuses on four research questions:

\begin{description}[align=left, left=10pt, itemsep=1ex]
\item [\textbf{\RqOne{}:}] \textbf{\RqOneText{}}

Prior work has shown that a classifier trained on code-stylometry features can distinguish among different human developers. We conjecture that a classifier can be built to differentiate code developed by humans from code generated by AI tools.

\item [\textbf{\RqTwo{}:}] \textbf{\RqTwoText{}}

Coding styles are deemed \textit{gameable} if they can be easily and strategically altered or avoided with minimal effort, particularly to mask the AI-generated nature of the code.  Examples of such gameable styles include the use of empty lines and whitespace for readability, which can be quickly adjusted without significantly impacting the overall code structure or functionality. If a classifier relies heavily on gameable features, then it may be relatively easy to disguise a \aig solution through simple code edits. We hypothesize that a classifier built using non-gameable code-stylometry features can effectively identify \aig code.

\item [\textbf{\RqThree{}:}] \textbf{\RqThreeText{}}
    
We hypothesize that a dataset containing incorrect \aig solutions may exhibit distinctive characteristics that could potentially enhance a classifier's ability to identify \aig solutions.

\item [\textbf{\RqFour{}:}]\textbf{\RqFourText{}}

We hypothesize that as the complexity of coding problems increases, the discriminative features between human and \aig code may become more pronounced. This could be attributed to the unique problem-solving approaches employed by human developers compared to AI systems when faced with complex programming tasks.
\end{description}
(3) Our dataset comprising \numProb problems and \totalNumSol human and \aig solutions is itself a contribution that we make publicly available to other researchers working on the problem of identifying \aig code. We also provide a subset of the dataset comprising \numProbCorrect problems whose solutions have been checked for correctness.

We provide a replication package,\footnote{\url{https://zenodo.org/doi/10.5281/zenodo.10152237}} which includes raw data, feature lists, and code scripts.

\section{Motivation and Related work}

In this section, we provide a real-world example of code-style differences between human and \aig code to motivate our study and related work.

\subsection{Motivation}
Figure~\ref{fig:motivating_example} presents two Python programs, both of which compute the sum of all palindromic numbers within an input range of integers\footnote{\url{https://www.codechef.com/problems/SPALNUM}} Figure~\ref{fig:motivating_example_a} represents a user submission obtained from CodeChef, whereas Figure~\ref{fig:motivating_example_b} was generated by ChatGPT. The two solutions exhibit clear differences in coding styles. The AI-generated code includes empty lines, and helper functions, whereas the human-authored code uses shorter identifiers. 
These differences and other code-style patterns (e.g., frequency of different keywords, complexity of expressions) used in prior studies on author 
attribution led us to question whether we could use code-stylometry features to build a classifier that distinguishes between human-authored code and \aig code.

\begin{figure}
    \centering
        \begin{subfigure}{0.48\textwidth}
            \begin{lstlisting}[language=python,breaklines=true]
for _ in range(int(input())):
l, r = map(int, input().split())
result = 0
for i in range(l, r + 1):
    if str(i) == str(i)[::-1]:
        result += i
print(result)\end{lstlisting}
            \caption{Example of human code}
            \label{fig:motivating_example_a}
        \end{subfigure}
        \begin{subfigure}{0.48\textwidth}
            \begin{lstlisting}[language=python,breaklines=true]
def is_palindrome(n):
    return str(n) == str(n)[::-1]

def palindromic_numbers_sum(l, r):
    total = 0
    for n in range(l, r+1):
        if is_palindrome(n):
            total += n
    return total

t = int(input())

for i in range(t):
    l, r = map(int, input().split())
    result = palindromic_numbers_sum(l, r)
    print(result)\end{lstlisting}
            \caption{Example of ChatGPT code}
            \label{fig:motivating_example_b}
        \end{subfigure}
    \caption{Solutions to a palindrome problem from CodeChef}
    \label{fig:motivating_example}
\end{figure}

\subsection{Related Work}
\subsubsection{Detecting AI-generated Code}
The prevalence and potency of AI assistants have led researchers to start investigating the problem of detecting code generated by AI assistants. Puryear and Sprint~\cite{puryear_copilot_in_class_2022} investigated how well-established plagiarism detection tools, MOSS \cite{Aiken_moss}, Codequiry \cite{Codequiry}, and CopyLeaks \cite{Copyleaks}, could detect Copilot-generated solutions 

among a set of data science programming assignments. They found that Copilot-generated solutions exhibited little similarity to solutions authored by students. The highest observed similarity, identified by MOSS at 36\%, fell well below the thresholds of similarity between student solutions that suggest plagiarism. 
Moreover, when ``similar'' Copilot and student solutions were manually inspected, the researchers determined that code similarities often reflected standard, commonly employed coding solutions or expected variable declarations. 
In work that is closest to ours, Bukhari et al.~\cite{bukhari2023distinguishing} attempt to use machine learning to distinguish between 28 student-authored and 30 AI-generated solutions for a C-language programming assignment involving singly-linked lists. Their approach leverages lexical and syntactic features in conjunction with multiple machine-learning models, achieving an accuracy rate of 92\%. 

We are also starting to see commercial tools such as HackerRank \cite{hackerrank2021} and Coderbyte \cite{coderbyte2021} that claim to identify AI-generated code within user-submitted code. Unfortunately, evidence of their performance has not been provided and is not freely available for third-party evaluation. 

\begin{figure*}[h]
  \centering
  \includegraphics[width=\linewidth]{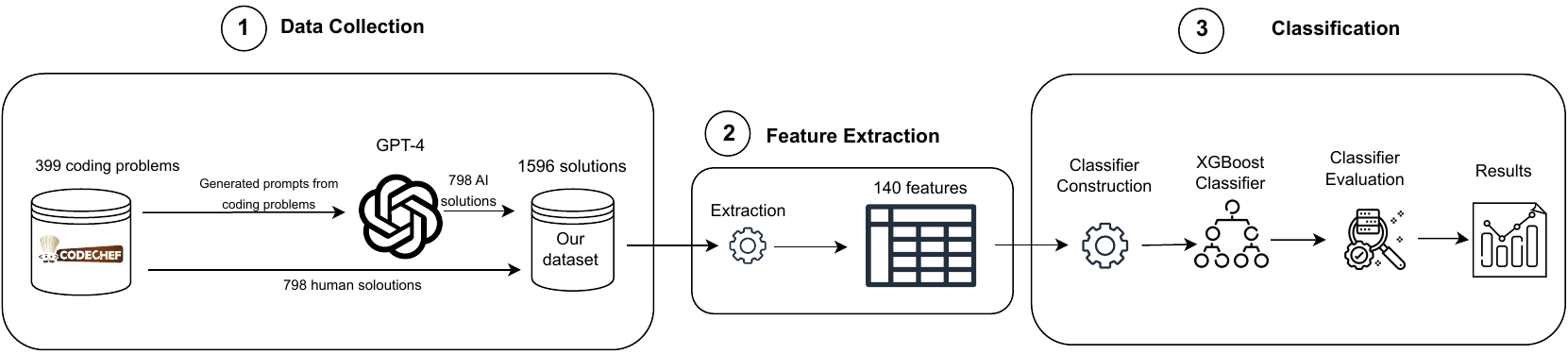}
  \caption{An overview of our approach in detecting \aig code}
  \Description{An overview of our approach in detecting \aig code}
  \label{fig:study_overview}
\end{figure*}

Our study expands on the body of work in this emerging field, employing a more diverse problem set and more descriptive features for interpretability compared to the study by Bukhari et al.~\cite{bukhari2023distinguishing}.

\subsubsection{Code Stylometry} 
A related research problem focuses on identifying code authorship, typically by using code stylometry, which analyzes distinct coding styles that reflect patterns in the way programmers write code. There exists a substantial body of work on coding constructs that can serve as distinctive identifiers of individual coding styles. Pioneering work by Oman and Cook in 1989 \cite{oman_programming_1989} analyzed the authorship of 18 distinct Pascal programs published in six independently authored computer science textbooks. More recent studies have used code stylometry for authorship attribution \cite{caliskan-islam_-anonymizing,  alsulami_source_2017, dauber2018gitblame, ullah_programmers_2020, frankel2021mlaa, hao_towards_2022} and plagiarism detection \cite{opgen-rhein_requirements_2019, dong_novel_2020, viuginov_machine_2021}. Although these terms are sometimes used interchangeably, authorship attribution deals with identifying the author of code, while plagiarism detection, on the other hand, assumes the author is known and aims to identify instances of unoriginal code \cite{kalgutkar_code_2019}. 
In these studies, different code stylometry features were found to be effective,
with layout or typographical features~\cite{oman_programming_1989} proving to be more accurate than Halstead’s metrics~\cite{halstead_natural_1972},\footnote{Halstead's metrics are measurable properties based on the author's hypothesis that the structure of code is based on two independent properties---operators and operands.}  a conventional complexity metric. Some studies used syntactic features~\cite{alsulami_source_2017, dauber2018gitblame, ullah_programmers_2020, frankel2021mlaa, hao_towards_2022}, whereas others combined layout, lexical, and syntactic features~\cite{caliskan-islam_-anonymizing, dong_novel_2020, viuginov_machine_2021}.

Our work leverages many of the code stylometry features used in the above studies for a new purpose: to distinguish between human-authored and \aig code.

\subsubsection{Machine Learning Approaches}
More generally, machine learning techniques have been applied in various code analysis tasks such as testing \cite{nair_leveraging_2019}, defect defection \cite{aggarwal_software_2020}, refactoring \cite{tollin_change_2017, kumar_method_2019}, vulnerability detection \cite{kronjee_discovering_2018, kaur_detecting_2018, eshghie_dynamic_2021}, program comprehension \cite{sun_open_source_lrning_2019}, code smells detection \cite{peiris_towards_2014}, authorship attribution \cite{caliskan-islam_-anonymizing,  alsulami_source_2017, dauber2018gitblame, ullah_programmers_2020, frankel2021mlaa, hao_towards_2022}, and plagiarism detection \cite{opgen-rhein_requirements_2019, dong_novel_2020, viuginov_machine_2021}. In our work, we use a machine learning technique to determine whether or not a program is \aig.

\section{Study Design}
In this section, we describe the different phases of our approach, including data collection, feature extraction, and classification. These phases are illustrated in Figure \ref{fig:study_overview}.

\subsection{Data collection}
The data collection phase is depicted on the left of Figure~\ref{fig:study_overview}. We collected Python problems and human solutions from a repository of programming problems, and we used an AI assistant to generate AI solutions. 
We chose Python specifically due to its status as a beginner-friendly language~\cite{briggs2012python} commonly adopted in introductory programming courses~\cite{kunkle2016introcourse}. To ensure a wide range of programming problems, we chose CodeChef as our problem repository. CodeChef is a renowned competitive coding platform known for offering problems of varying difficulty levels. We believe that there are no inherent differences across other competitive programming platforms, so our approach's performance on CodeChef should be similar to its performance on these other platforms.

Due to the absence of public APIs, we extracted data by scraping CodeChef's website, obtaining both problem sets and user submissions. The data collection process is divided into two steps, as highlighted in Figure~\ref{fig:study_overview}. We briefly describe each step below. \\
\textbf{Problem Set and Human Solutions Extraction.} In this step, we curated a problem set of coding problems from CodeChef and their corresponding human solutions.  

\begin{table}[t]
  \caption{Difficulty Levels of Selected CodeChef Problems}
  \label{tab:problems_diff_level}
  \begin{tabular}{rcc}
    \toprule
    Level       &   Range        &  Count   \\
    \midrule
    Beginner    &   0 - 999       & 12      \\
    1* Beginner &   1000 - 1199   & 45      \\
    1* Advanced &   1200 - 1399   & 71      \\
    2* Beginner &   1400 - 1499   & 55      \\
    2* Advanced &   1500 - 1599   & 56      \\
    3* Beginner &   1600 - 1699   & 60      \\
    3* Advanced &   1700 - 1799   & 53      \\
    4*          &   1800 - 1999   & 30      \\
    5*          &   2000 - 2199   & 14      \\
    6*          &   2200 - 2499   &  2      \\
    7*          &   2500 - 5000   &  1      \\
  \bottomrule
                &                 & 399
\end{tabular}
\end{table}

%@=[]=====>
CodeChef assigns each problem on its platform a difficulty score and classifies ranges of difficulties into 11 buckets, as depicted in Table~\ref{tab:problems_diff_level}. To ensure that our study's problem set has a good distribution with respect to difficulty, we fetched the 100 most popular problems from each difficulty level. Here popularity refers to the number of accepted solution attempts that existed when the data was scraped (November 2023). In the case that a user submits multiple solutions to the same problem statement, we collect only the latest correct solution submitted by the user. Thus, we began our filtering process with 1100 problem statements. 

To further refine our problem set, we selected those problems with at least two correct solutions submitted in 2020. This year was specifically chosen as it postdates the end-of-life for Python 2 in January 2020 and predates the release of AI assistants like Copilot in October 2021 and ChatGPT in November 2022. We purposefully excluded solutions submitted after the release of these AI assistants to ensure that the human-authored solutions in our dataset are not polluted with AI-generated code.
We also purposefully excluded incorrect solutions to avoid including incomplete submissions that do not reflect a typical attempt at a successful solution. In the context of our study, a solution is deemed \textit{correct} if it passes CodeChef's public tests.\footnote{In the context of competitive programming, a submission is deemed \textit{correct} if it passes all public and private test cases; however, we had access only to CodeChef's public tests for each problem.} Such a solution represents successfully interpreted and executed code, effectively solving the specified problem as far as the public tests are concerned.

After this filtering, the number of problems in our dataset was reduced to 419. Upon closer inspection, we found 20 problems that were tagged as ``Python3'' problems but had only Python2 solutions. These problems were therefore excluded from our study, resulting in a problem set comprising \numProb problems, each with at least two correct human-authored Python3 solutions submitted in 2020. In addition, each problem includes comprehensive details from the platform such as the problem statement, unique problem code ID, input and output formats, assigned difficulty score, subtasks, constraints, problem names, user-assigned tags, computed tags, and sample test cases containing input, output, and explanations.

\begin{table}[t]
  \caption{Final Problem Set Binned into 3 Classes of Difficulty }
  \label{tab:binned_problems_diff_level}
  \begin{tabular}{rccc}
    \toprule
    \textbf{Difficulty}       &   \textbf{Difficulty Scores - Range}        &  \textbf{Average} &  \textbf{Count}  \\
    \midrule
    Easy   &   828 - 1417    & 1224.95 & 133      \\
    Medium &   1419 - 1646   & 1529.51 & 133     \\
    Hard &   1647 - 3420   & 1827.50 & 133     \\
  \bottomrule
\end{tabular}
\end{table}

After filtering we noted that we were left with very few problems in the highest levels of difficulty as shown in Table~\ref{tab:problems_diff_level}. This could be attributed to the fact that competitive programmers often choose C/C++ over Python for various reasons, particularly in contests involving problems of higher difficulty. 
To mitigate issues with inference due to disparities in the number of problems per category, we re-binned the \numProb problems into three classes of difficulty (easy, medium, and hard) of equal size. The classes contain 133 problems each, with average difficulty scores of 1224.95, 1529.51, and 1827.50 respectively as shown in Table~\ref{tab:binned_problems_diff_level}. This reclassification of difficulty attempts to ensure a balanced representation across categories and enables a fair evaluation of how well our classifiers fare with respect to problems of different difficulty.

\noindent\textbf{Not including comments.} Although comment-based features like commentsDensity~\cite{caliskan-islam_-anonymizing}, inlineCommentsDensity~\cite{oman_programming_1989}, and blockedCommentsDensity~\cite{oman_programming_1989} have been used in past works on authorship attribution, we have chosen not to include comments for two reasons:
\begin{enumerate}
    \item Comments may make it too easy for a classifier to determine if the code is human-authored or \aig. This is because when you specifically ask for comments from a model like \ai, the number of comments is far more than any human would normally write. By excluding comments we handicap our approach and thus provide a lower bound for our classifier. Besides this, comment-based features are easily gameable. 

    \item The amount of comments present in \aig code varies by the prompt that we give. If we simply query the API of the model to write a program for the problem at hand, the model generates code only and no comments. Alternatively, if we prompt the model to explain the code, then almost every line of code is commented on. Given that the focus of this work is to explore whether we can differentiate between human-authored and \aig code, we did not want the choice of prompt to be a variable of the experiment.
\end{enumerate}

Hence we explicitly asked the \ai model not to explain the code and then we stripped any comments that may have been included even by mistake. We apply the same comment-stripping technique to remove any comments from the human code as well. Therefore in this study, we explicitly avoid using comment-related features for classification. 

\begin{figure}[t]
\centering
\begin{minipage}{\linewidth}

\hrulefill

You are an expert Python Programmer. Your job is to look at a programming puzzle provided by the user and output 2 different ways to solve the solution in python.

The Input is provided with the following contents:

\{The problem statement\}

\{How the input would be formatted\},

\{Format to be followed in the output generated\},

\{Constraints on the variables specified in the problem\}

Make sure to take the input from the user considering the input format
Output should be printed as defined in the output format

Do not attempt to explain the solution only output the code in the following format:

[PYTHON1]

\{Solution to given puzzle in Python\}

[$\backslash$PYTHON1]

[PYTHON2]

\{Alternate solution to given puzzle in Python\}

[$\backslash$PYTHON2]

\hrulefill

\caption{Prompt used for generating 2 AI code solutions}
\label{fig:generation_prompt}
\end{minipage}
\end{figure}

\noindent\textbf{AI Solutions Generation.} To generate AI solutions to the problems in our problem set we used \ai (Version 0613), which is one of the most powerful and easily accessible generative models available to consumers as of November 2023 through OpenAI \cite{jiao2023evaluating}. We set the temperature to 0 so that our results are reproducible (at the time of this writing, setting a seed for consistent generation was not available through the API). We used the prompt shown in Figure~\ref{fig:generation_prompt} to obtain two \ai solutions for each of the \numProb problems in our problem set, resulting in \numSol unique solutions. 
In constructing our prompt, we employed strategies recommended by OpenAI for effective prompt engineering.\footnote{\url{https://platform.openai.com/docs/guides/prompt-engineering/strategy-provide-reference-text}} The specific strategies we followed are outlined below:
\begin{itemize}
\item \textbf{Include Details in Your Query to Get More Relevant Answers:} The prompt specifies the details of the task by defining the format of the input (problem statement, input format, output format, constraints) and the required output (two Python solutions). This helps in getting relevant and specific answers tailored to the given programming problem.

\item \textbf{Ask the Model to Adopt a Persona:} The prompt begins with "You are an expert Python Programmer." This tactic of persona adoption sets a context for the responses expected and guides the AI to frame its responses within the expertise of a Python programmer.

\item \textbf{Use Delimiters to Clearly Indicate Distinct Parts of the Input:} The prompt uses a structured format with clear delimiters, such as [PYTHON1] and [PYTHON2], to separate the two different solutions. This helps the AI understand that two distinct solutions are required and organizes the output in a clear, readable manner. It also enables us to programmatically process the outputs generated.

\item \textbf{Specify the Steps Required to Complete a Task:} While the prompt implicitly suggests the steps (understand the problem, code the solution), it doesn't explicitly break down the task into smaller steps. In tasks like programming, outlining steps such as analyzing the problem, considering algorithms, and then coding can enhance the quality of the response.
\end{itemize}

We opted for a zero-shot inference approach. We intentionally did not constrain the output length within the prompt or provide a detailed step-by-step breakdown,  among the other suggested strategies to accommodate the diverse nature of problems in our dataset. While this prompt could be further refined, our goal was to develop a pragmatic, 'best effort' prompt reflective of what a typical user might employ.

All generated \ai solutions were syntactically valid and could be successfully parsed, which was important for AST-based features. In the case of duplicate solutions to a problem, we reran the prompt to obtain a new solution to swap in. 

Private tests for the problems could not be scraped from the platform, thus we evaluated the \aig solutions on the available public tests to check whether they were correct to some degree. We found that only 137 problems had two \ai solutions that satisfy the available public tests, and another 24 problems had only one of the solutions passing the test cases. We used this information to create a sanitized set of \numProbCorrect problems that includes correct \ai solutions for each problem and an equal number of unique and correct human solutions to those problems for \RqThree.

\begin{table*}[t]
\caption{Code Stylometry and Code Complexity Features} 
\label{tab:features}
\resizebox{\textwidth}{!}{%
\begin{tabular}{l p{16cm}}
\toprule
\textbf{Feature} & \textbf{Description}\\
\midrule
ASTNodeTypesTF \cite{caliskan-islam_-anonymizing} & Term frequency of 130 possible AST node types excluding leaves\\
ASTNodeTypeAvgDep \cite{caliskan-islam_-anonymizing} & Average depth of 130 possible AST node types excluding leaves.\\
avgFunctionLength \cite{faidhi_plagiarism_1987} & The average length of lines in a function.\\
avgIdentifierLength  \cite{faidhi_plagiarism_1987} & The average length of identifier names.\\
avgLineLength \cite{caliskan-islam_-anonymizing} & The average length of characters in each line.\\
avgParams \cite{caliskan-islam_-anonymizing} & The average number of parameters across all functions.\\
branchingFactor \cite{caliskan-islam_-anonymizing} & Average branching factor of the code's AST.\\
% commentsDensity \cite{caliskan-islam_-anonymizing} & The number of comments divided by source code lines.\\
cyclomaticComplexity \cite{mccabe_complexity_1976} & The number of decisions within a block of code.\\
emptyLinesDensity~\cite{caliskan-islam_-anonymizing} & The number of empty lines divided by source code lines.\\
keywordsDensity \cite{caliskan-islam_-anonymizing} & Frequency of Python keywords divided by source lines of code.\\
maintainabilityIndex \cite{coleman_maintainability_1994} & A metric that gauges the ease of supporting and modifying the source code.\\
maxDecisionTokens & The maximum number of tokens in decision conditions excluding ternary conditions.\\
maxDepthASTNode \cite{caliskan-islam_-anonymizing} & Maximum depth of an AST node.\\
nestingDepth \cite{caliskan-islam_-anonymizing} & Deepest level to which conditional statements, loops, and functions are nested within each other.\\
numAssignmentStmtDensity \cite{donaldson_plagiarism_1981} & The total number of assignment statements divided by source code lines.\\
numClassesDensity & The total number of classes divided by source code lines.\\
numFunctionCallsDensity \cite{donaldson_plagiarism_1981} & The total number of function calls divided by source code lines.\\
numFunctionsDensity \cite{donaldson_plagiarism_1981} & The number of functions divided by source code lines.\\
numInputStmtsDensity \cite{donaldson_plagiarism_1981} & The total number of input statements divided by source code lines.\\
numKeywordsDensity \cite{caliskan-islam_-anonymizing} & The total number of unique Python keywords divided by source code lines.\\
numLiteralsDensity \cite{caliskan-islam_-anonymizing} & The number of literals divided by sloc.\\
numStatementsDensity \cite{donaldson_plagiarism_1981} & The total number of statements divided by source code lines.\\
numVariablesDensity \cite{donaldson_plagiarism_1981} & The total number of assignment variables divided by source code lines.\\
numberOfDistinctOperands \cite{halstead_natural_1972} & The number of distinct operands.\\
numberOfDistinctOperators \cite{halstead_natural_1972} & The number of distinct operators.\\
sloc  \cite{grier_tool_1981} & The total number of source code lines.\\
stdDevLineLength  \cite{caliskan-islam_-anonymizing} & The standard deviation of character lengths of each line.\\
stdDevNumParams \cite{caliskan-islam_-anonymizing} & The standard deviation of the number of parameters across all functions.\\
totalNumberOfOperands \cite{halstead_natural_1972} & The total number of operands.\\
totalNumberOfOperators \cite{halstead_natural_1972} & The total number of operators.\\
whiteSpaceRatio \cite{caliskan-islam_-anonymizing} & The ratio of whitespace characters to non-whitespace characters.\\
\bottomrule
\end{tabular}
}
\end{table*}

\subsection{Feature Extraction}
Our study leverages a combination of layout, syntactic, and lexical features that have been effectively used in previous studies for authorship attribution and plagiarism detection among human programmers~\cite{caliskan-islam_-anonymizing, faidhi_plagiarism_1987, donaldson_plagiarism_1981, grier_tool_1981}. Layout features refer to the visual organization of code, such as indentation and spacing. Lexical features, on the other hand, are derived from analyzing the tokens within the code, capturing elements such as keywords and literals. While syntactic features are extracted based on the code's structural patterns, involving the arrangement and relationships between various code elements. Our study also incorporates Halstead's metrics~\cite{halstead_natural_1972}, which have been used in previous studies for authorship attribution~\cite{berghel_1984_similarity_measurement, oman_programming_1989}. We also included additional complexity metrics such as maintainability index~\cite{coleman_maintainability_1994} and cyclomatic complexity~\cite{mccabe_complexity_1976} to enrich our approach.

In the feature extraction phase (shown in the middle of Figure~\ref{fig:study_overview}), we iterate through the Python solution files, systematically generating these code stylometry and complexity features essential for training and evaluating our classifier.

We extracted 31 base features (shown in Table~\ref{tab:features}) plus variants, leading to \numFeatures features. Most base features have no variants. Feature keywordsDensity has 28 variants, out of 35 Python keywords; these are listed in the replication package. Features ASTNodeTypesTF and ASTNodeTypeAvgDep each have 42 variants, out of 130 possible AST node types; these are listed in the replication package. The prefixes ``nttf\_'' for ASTNodeTypesTF, ``ntad\_'' for ASTNodeTypeAvgDep, and the suffix ``\_Density'' for KeywordsDensity were adopted to correlate variants with their respective base features. 

Our approach to feature extraction varies somewhat from prior studies~\cite{caliskan-islam_-anonymizing, donaldson_plagiarism_1981,  oman_programming_1989}. We normalized our features by source lines of code rather than by character count~\cite{caliskan-islam_-anonymizing} or by omitting normalization altogether~\cite{donaldson_plagiarism_1981,  oman_programming_1989}. Additionally, we refrained from logarithmic transformations of some features, as practiced by Caliskan et al. ~\cite{caliskan-islam_-anonymizing}, to facilitate the ease of interpretability of our feature set, particularly for visual analysis. For the nestingDepth feature, we considered node types rather than actual tokens. The MaintainabilityIndex feature measures code maintainability by evaluating complexity and modularity and is calculated using complexity metrics such as Cyclomatic Complexity, and SLOC. Our version 
was computed with the Radon Python library,~\footnote{\url{https://pypi.org/project/radon/}} which uses a modified formula different from the study by Coleman et al. ~\cite{coleman_maintainability_1994}. 
Although our dataset contains only single-file solutions, the MaintainabilityIndex feature is included in our study as it may yield insights into the relative maintainability of code produced by AI assistants compared to code authored by humans, potentially impacting the performance of our classifier.

Of the \numFeatures features extracted, four are Halstead metrics, selected to explore their viability in this context. Their ineffectiveness in the context of authorship attribution in human code has been pointed out by Berghel and Sallach \cite{berghel_1984_similarity_measurement} and Oman and Cook~\cite{oman_programming_1989}, but we included them for completeness. Our research is directed at assessing how well these metrics can identify \aig code.
%amidst such concerns.

\subsection{Classification}
The righthand side of Figure \ref{fig:study_overview}  provides an overview of the classification phase, where we construct a model to distinguish between \aig and human-authored code and subsequently evaluate its performance. We describe each step of this phase below. \\
\textbf{Classifier Construction.}
To construct our classifier, we chose XGBoost~\cite{xgboost_2016}, because it is an effective and scalable machine learning algorithm. Additionally, in the study by Bukhari et al~\cite{bukhari2023distinguishing}, XGBoost with syntactic and lexical features had the best performance when considering accuracy and F1 score. This allows us to compare our approach with the best of the earlier study's approaches.
XGBoost constructs decision trees iteratively, refining the model by correcting misclassifications at each step. During training, the algorithm optimizes an objective function to strike a balance between prediction accuracy and model simplicity. At each tree, the algorithm assigns scores to examples, and each example's final prediction is calculated by summing the scores~\cite{xgboost_2016}. Through this aggregation process, the resulting model classifies code as either human-authored or \aig. \\
\textbf{Classifier Evaluation.} To evaluate the classifier's performance on unseen data, we employ ten-fold cross-validation, which divides the dataset into ten subsets. This approach provides robust assessments by training on nine subsets and testing on the remaining subset. This process is iterated ten times, guaranteeing that each subset serves as the test set exactly once. 

To enhance the evaluation process and avoid data poisoning during training, we grouped solutions based on the specific coding problem they addressed, resulting in \numProb distinct groups corresponding to the \numProb coding problems in our dataset. We employed GroupKFold to ensure that each group, representing solutions to a particular problem, appeared only once in the test set across all folds. This grouping strategy maintains the integrity of the evaluation by preserving the context of solutions within each problem, preventing the model from training and testing on the same problem sets since each problem set contains multiple solutions.
% making decisions based solely on the similarity of solutions due to shared coding requirements.

The model's performance is measured in terms of accuracy, recall, precision, F-measure, and AUC-ROC. 

\textbf{Baselines Comparison.}
In our model comparison, we evaluate our approach alongside two baselines: (1) a naive baseline approach, based on the assumption that \aig code can be detected through random guessing, and (2) the approach presented by Bukhari et al. \cite{bukhari2023distinguishing} that identified AI-generated solutions for C programming assignments. To benchmark our classifier, we replicated the methodology of Bukhari et al. \cite{bukhari2023distinguishing} using Python and an XGBoost classifier, motivated by XGBoost performing the best in their study on syntactic and lexical features. We evaluate the performance of our classifier using metrics such as accuracy, recall, precision, F-measure, and AUC-ROC.

% Given that our case study focuses solely on the CodeChef dataset, we did not conduct statistical significance tests over the second baseline. Such tests typically require multiple datasets to validate a hypothesis, which our study does not include. Additionally, while a probability-based method was considered, the risk of high probability misclassifications undermines the accuracy of this method. 
% We also contemplated the use of ten-fold cross-validation for our analysis. However, this approach has limitations, as it provides only ten data points and is influenced by how the folds are selected. Future research could expand on this study by utilizing multiple datasets, enabling effective statistical significance tests.

Given that our case study focuses solely on utilizing the CodeChef dataset, we will not employ statistical significance tests like the Mann-Whitney U or T-test between our approach and the second baseline. These tests typically require multiple datasets, a requirement our single-dataset study does not meet. Additionally, such tests cannot be employed on the classifier’s raw prediction scores due to the presence of high probability misclassifications. Furthermore, applying statistical significance tests to performance metrics derived from ten-fold cross-validation is problematic due to the overlap of the training dataset across folds, with about 80\% of data shared between each pair of training sets \cite{dietterich1998approximate}. This overlap leads to interdependence among the folds, thereby violating the independence assumption of these tests. As a result, we risk incorrectly rejecting the null hypothesis even in instances where there may be no actual difference.

\section{Results}
We present the results of our model evaluation in relation to our research questions. For each research question, we first outline our approach to addressing it, followed by the observed results. \\
\emph{\textbf{(\RqOne{}.)} \RqOneText{}}

\textbf{\underline{Approach.}} To address our research question, we build a classifier trained on code stylometry features. After training, we evaluate its performance using ten-fold cross-validation, focusing on metrics such as precision, recall, F-measure, and AUC-ROC. High scores in these metrics will support our research hypothesis, demonstrating our approach's ability to distinguish human-authored code from AI-generated code.
Additionally, we compare this classifier with an alternative classifier trained on the same feature set but augmented with Halstead’s metrics. The objective is to evaluate the impact of Halstead’s metrics in detecting \aig code. This will involve a comparative analysis of the classifiers' performances with and without Halstead’s metrics. Moreover, we contrast the performance of our classifier with two baselines.

The Naive Baseline is a random guess and its performance metrics can be computed by applying statistics to our dataset. The precision is calculated by dividing the number of \aig code by the total number of solutions:

    \begin{gather}
    \text{Precision} = \frac{\text{number of \aig code}}{\text{total number of solutions}} = 0.5
    \end{gather}

The recall is 0.5, reflecting the classifier's two possible outcomes---identifying code as either GPT-4 generated or human-authored. This results in a probability of 0.5 for classifying solutions as \aig. Based on precision and recall values, we compute the $F_1$ score of the naive baseline as:

    \begin{gather}
    F_1 = 2 \times \frac{Precision \times Recall}{Precision + Recall} = 0.5
    \end{gather}\\

\begin{table}[t]
\centering
\caption{Classifier Performance Comparison among Different Approaches for Distinguishing between AI-generated and Human-authored Code}
\label{tab:classifier_performance}
\begin{tabular}{l c c c c c}
\toprule
 & \multicolumn{2}{c}{\textbf{Our Approach}} & \multicolumn{3}{c}{\textbf{Baseline}} \\
\cmidrule(lr){2-3} \cmidrule(lr){4-6}
& \textbf{All} & \textbf{Non-Gameable} & \textbf{Naive} & \multicolumn{2}{c}{\textbf{n-grams + L}} \\
& & & & \textbf{n = 2} & \textbf{n = 3} \\
\midrule
 Accuracy  &  0.91 &  0.89 &  -    &  0.86 &  0.88 \\
 Precision &  0.91 &  0.89 &  0.5  &  0.86 &  0.87 \\
 Recall    &  0.91 &  0.89 &  0.5  &  0.88 &  0.88 \\
 F1-score  &  0.91 &  0.89 &  0.5  &  0.87 &  0.88 \\
 AUC-ROC   &  0.91 &  0.89 &  -    &  0.86 &  0.88 \\

\bottomrule

\end{tabular}
\end{table}
    
\textbf{\underline{Results.}} As shown in Table \ref{tab:classifier_performance},\footnote{It is just a coincidence that the performance metrics for our models all have the same value when considering all features (0.91) and all non-gameable features (0.89)} our approach achieved a high average precision and recall of 0.91, ensuring accurate and comprehensive identification of \aig code. This highlights the potential of code stylometry in differentiating between \aig and human-authored code. Comparing the classifiers, one with and the other without Halstead metrics, we found a striking similarity in their performance metrics. Between both classifiers, all the metrics considered were the same, except recall which was higher for the classifier with Halstead metrics by 0.01. This observation suggests that the presence of Halstead metrics does not considerably enhance the classifier's ability to distinguish between human-authored and \aig code supporting past work~\cite{grier_tool_1981, leach_1995_eval_student_programs}. 

When compared to the baselines, our classifier shows a considerable improvement. It notably outperformed the Naive Baseline, which has a precision and recall of 0.5, demonstrating that our classifier considerably exceeds what would essentially be random guessing. We also compare with the work presented in  Bukhari et al. \cite{bukhari2023distinguishing} that incorporated lexical features of 2-4 n-grams. However, in our replication, we could only process 2-3 n-grams due to the memory-intensive nature of the task. The data for the 4-gram model was at least 212.33 GBs, resulting in out-of-memory errors in our machine (Macbook Air 2020, with 16GBs of RAM). Also, it took more than 8 hours to extract the data for the 4-gram model. However, our classifier has no such issue. We were also able to achieve higher precision and recall by 4\% and 3\% respectively.

Our study demonstrates a higher predictive power over the n-grams baseline and provides interpretable predictions. Unlike its n-gram-based features, which may obscure the reasoning behind predictions, our model uses a feature set that clarifies the decision-making process. To understand the influence of specific features on our model’s predictions, we used the SHAP framework~\cite{lundberg_2017_SHAP}, a method renowned for its interpretability of machine learning models. SHAP offers tools for both local (individual) and global (overall) explanations of model predictions.

\begin{figure}[t]
  \centering
  \includegraphics[width=\linewidth]{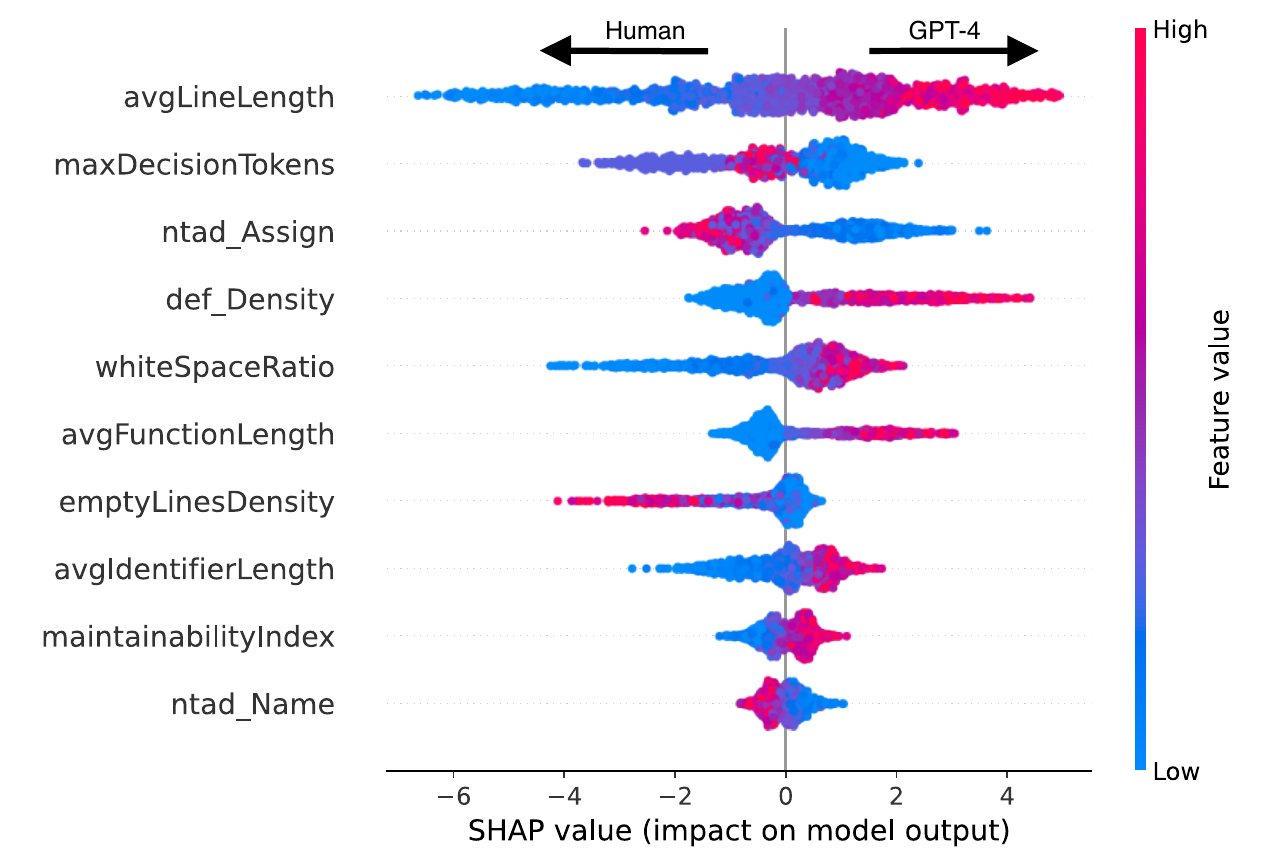}
  \caption{SHAP feature importance of our approach}
  \Description{SHAP feature importance of our approach}
  \label{fig:summary_plot_oa}
\end{figure}

The global interpretive power of our classifier is demonstrated in the SHAP summary plot depicted in Figure~\ref{fig:summary_plot_oa}. This plot visualizes key features in our classifier, arranging them on the y-axis by their aggregate SHAP values, with the highest at the top. The x-axis displays these SHAP values, showing how each feature shifts the prediction from a neutral base value, indicated by the vertical line at 0 on the x-axis. Deviations to the left or right increase the likelihood of the prediction being a human or \ai class, respectively. The plot uses a blue-to-red color gradient to signify feature magnitudes, and the data points represent feature values across instances. 

As depicted in Figure~\ref{fig:summary_plot_oa}, the \textit{avgLineLength} is the most important feature, distinctly separating human and \ai classes where its lower values are typically associated with the human class, whereas higher values are associated with the \ai class. This implies that a line of code from a human is shorter on average compared to a line of code from \ai.
In contrast, the \textit{ntad\_Name} feature, is the tenth most important feature and quantifies the average depth at which the \textit{Name} node occurs within an AST. The \textit{Name} node in Python's AST represents identifiers, which are the names of variables, functions, classes, modules, or other objects in the code. The distribution of the \textit{ntad\_Name} feature across all instances shows a relatively narrow range of SHAP values, suggesting that its impact on the model is less compared to the features ranked above it. It is important to note that SHAP functions as an explanation model that provides an interpretable approximation of our classifier. Therefore, while SHAP values offer a simplified and interpretable view, the feature importance rankings derived directly from our classifier are based on the classifier's internal mechanisms. Consequently, the feature importance rankings from our classifier, do not entirely align with SHAP's. Despite these differences, there is considerable overlap in the primary features identified by both methods. This overlap highlights the value of SHAP in interpreting the model's predictions, offering insights into how features influence outcomes rather than detailing the classifier's internal mechanisms.

\emph{\textbf{(\RqTwo{}.)} \RqTwoText{}}

\textbf{\underline{Approach.}} To assess the impact of non-gameable features on the detection of \aig code, we build a classifier that excludes gameable features.

In our analysis, we consider the non-code layout features \textit{emptyLinesDensity} and \textit{whiteSpaceRatio} as gameable features. After training on the non-gameable features, we evaluate the performance of this classifier in contrast to our classifier with both gameable and non-gameable features. We also evaluate this classifier with the same n-grams baseline compared in \RqOne~as the baseline does not include the features we consider gameable in its feature set. Through these comparisons, we aim to evaluate the relative importance of using only non-gameable features in the classification of code as human-authored or \aig. 

\begin{figure}[t]
  \centering
  \includegraphics[width=\linewidth]{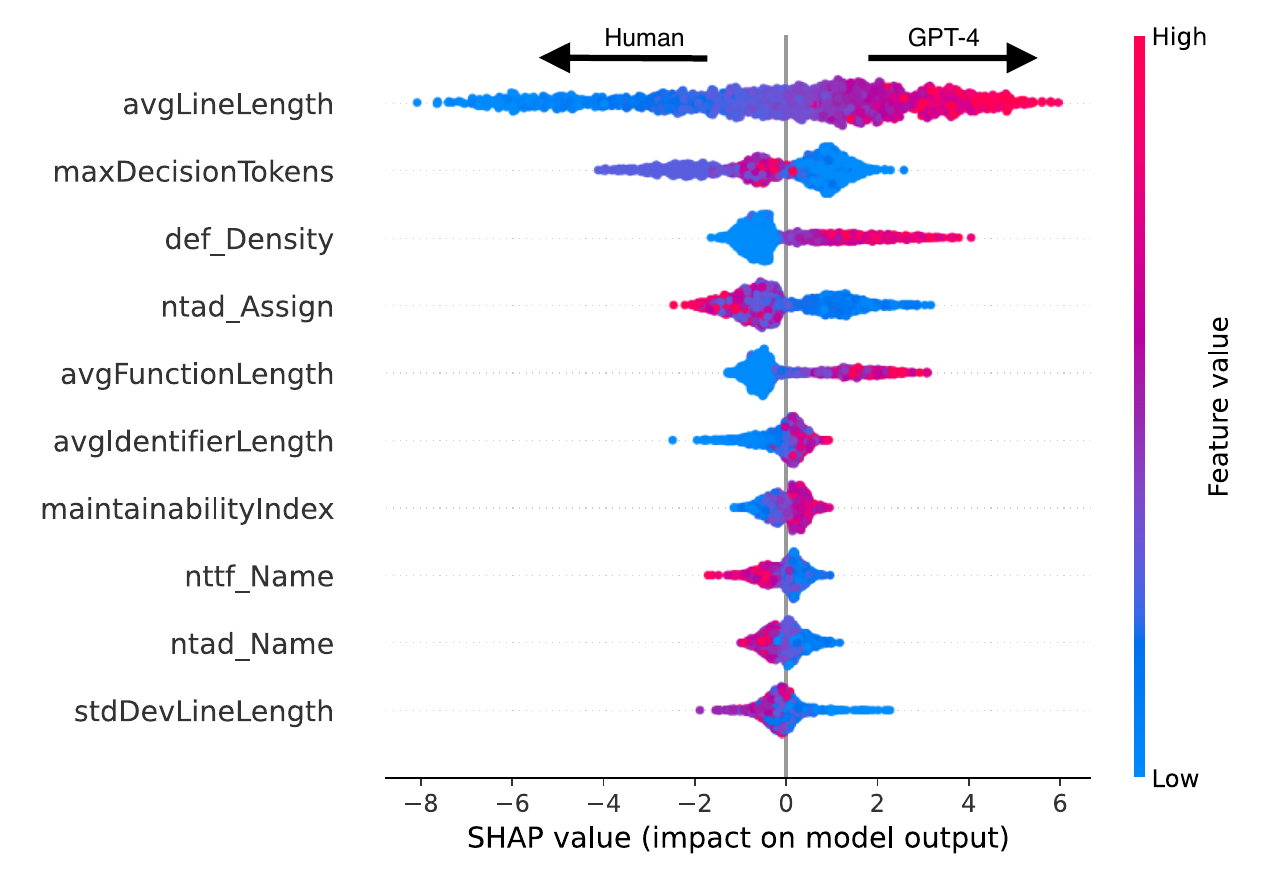}
  \caption{SHAP feature importance of non-gameable features}
  \Description{SHAP feature importance of non-gameable features.}
  \label{fig:summary_plot_non_gameable}
\end{figure}

\textbf{\underline{Results.}} As shown in Table
\ref{tab:classifier_performance}, there is a noticeable but not severe drop in performance for the non-gameable classifier compared to the classifier built on gameable and non-gameable features. This suggests that although gameable features contribute to the classifier's accuracy, non-gameable features alone still provide a high predictive power. In comparison to the n-grams baseline, the non-gameable classifier still performs better and is interpretable as evident in the summary plot of Figure~\ref{fig:summary_plot_non_gameable}. The plot reveals that aside from the two gameable features among the ten most important features of the classifier trained on both gameable and non-gameable features shown in Figure~\ref{fig:summary_plot_oa}, the other features remain consistent with a slight reordering. The absence of these two gameable features accounts for the performance dip in the non-gameable classifier. Consequently, \textit{stdDevLineLength} and \textit{nttf\_Name}, the latter representing the term frequency of the Name node, now appear in the top ten. The \textit{stdDevLineLength} feature influences predictions towards the \ai class at lower values and towards the human class at intermediate values. Conversely, \textit{nttf\_Name} influences predictions towards the human class at higher values and towards the \ai class when lower.

\begin{table}[t]
\centering
\caption{Classifier Performance Comparison on Correct and Randomly Sampled Solutions}
\label{tab:classifier_performance_correct_solutions}
\begin{tabular}{l c c c c c c}
\toprule
& \multicolumn{2}{c}{\textbf{Our Approach}} & \multicolumn{4}{c}{\textbf{Baseline (n-grams + L)}} \\
\cmidrule(lr){2-3} \cmidrule(lr){4-7}
& \multicolumn{1}{c}{\textbf{C}} & \multicolumn{1}{c}{\textbf{R}} & \multicolumn{2}{c}{\textbf{n = 2}} & \multicolumn{2}{c}{\textbf{n = 3}} \\
& & & \multicolumn{1}{c}{\textbf{C}} & \multicolumn{1}{c}{\textbf{R}} & \multicolumn{1}{c}{\textbf{C}} & \multicolumn{1}{c}{\textbf{R}} \\
\midrule
 Accuracy  & 0.86 & 0.87 & 0.83 & 0.84 & 0.87 & 0.86 \\
 Precision & 0.87 & 0.87 & 0.83 & 0.84 & 0.87 & 0.86 \\
 Recall    & 0.86 & 0.88 & 0.81 & 0.85 & 0.87 & 0.85 \\
 F1-score  & 0.86 & 0.87 & 0.82 & 0.84 & 0.87 & 0.86 \\
 AUC-ROC   & 0.86 & 0.87 & 0.83 & 0.84 & 0.87 & 0.86 \\
\bottomrule
\end{tabular}
\smallskip
\footnotesize \\
\textbf{C = Correct Solutions, R = Random Solutions, L = Lexical Features}
\end{table}

\emph{\textbf{(\RqThree{}.)} \RqThreeText{}}

\textbf{\underline{Approach.}} To address this research question, we refine our dataset to include only correct solutions, resulting in a balanced dataset of 596 correct solutions from both humans and \ai. We aim to evaluate our classifier's predictive power by eliminating the potential noise incorrect solutions might introduce. This ensures that the detection of \aig solutions is based on inherent characteristics of code, not the errors they might contain.

\textbf{\underline{Results.}} Table \ref{tab:classifier_performance_correct_solutions} shows a slight decline in performance when compared to the classifier in Table \ref{tab:classifier_performance}. The minor decline could stem from the correct solutions being a smaller subset. 
Hence, we randomly sample the solutions comprising both correct and incorrect solutions with the same distribution of difficulty levels. When compared with the random solutions classifier, both perform almost the same. This implies that genuine differences in coding styles between human and \aig code are being detected, rather than errors introduced by incorrect solutions. When compared to the baseline \cite{bukhari2023distinguishing}, for correct solutions, our classifier outperformed the 2-gram baseline but was marginally less effective than the 3-gram. The classifier with randomly sampled solutions showed noticeable improvement over the 2-gram baseline and was marginally better than the 3-gram.

\begin{table*}[t]
\centering
\caption{Classifier Performance Comparison Across Levels of Problem Difficulty}
\label{tab:classifier_performance_difficulty_levels}
\begin{tabular}{l c c c c c c c c c}
\toprule
& \multicolumn{3}{c}{\textbf{Our Approach}} & \multicolumn{6}{c}{\textbf{Baseline (n-grams + Lexical Features)}} \\
\cmidrule(lr){2-4} \cmidrule(lr){5-10}
& \textbf{Easy} & \textbf{Medium} & \textbf{Hard} & \multicolumn{3}{c}{\textbf{n = 2}} & \multicolumn{3}{c}{\textbf{n = 3}} \\
\cmidrule(lr){5-7} \cmidrule(lr){8-10}
& & & & \textbf{Easy} & \textbf{Medium} & \textbf{Hard} & \textbf{Easy} & \textbf{Medium} & \textbf{Hard} \\
\midrule
 Accuracy  & 0.89 & 0.89 & 0.87 & 0.87 & 0.79 & 0.80 & 0.89 & 0.86 & 0.80 \\
 Precision & 0.87 & 0.88 & 0.89 & 0.85 & 0.80 & 0.79 & 0.89 & 0.87 & 0.80 \\
 Recall    & 0.91 & 0.90 & 0.86 & 0.89 & 0.77 & 0.82 & 0.88 & 0.85 & 0.81 \\
 F1-score  & 0.89 & 0.89 & 0.87 & 0.87 & 0.79 & 0.80 & 0.89 & 0.86 & 0.80 \\
 AUC-ROC   & 0.89 & 0.89 & 0.87 & 0.87 & 0.79 & 0.80 & 0.89 & 0.86 &  0.80 \\
\bottomrule
\end{tabular}
\end{table*}

\emph{\textbf{(\RqFour{}.)} \RqFourText{}}

\textbf{\underline{Approach.}} To address this research question, we construct separate classifiers for the three problem difficulty levels outlined in Table~\ref{tab:binned_problems_diff_level}. By training and evaluating these classifiers independently, we evaluate their performance and the potential impact of problem complexity on the classifier's ability to correctly identify \aig code. This stratified analysis allows us to understand the nuances of classifier performance across problem difficulty levels.

\textbf{\underline{Results.}} Table \ref{tab:classifier_performance_difficulty_levels} shows a correlation between the classifier performance and the difficulty of the problems. The performance of classifiers of easy and medium problem difficulty are close as they have the same F1-score of 0.89. The performance of the classifier with hard problems had a minor drop in performance with an F1-score of 0.87. When compared to the baseline, there is no considerable difference in the baseline for easy questions. For both medium and hard questions, our classifiers perform better than both 2-gram and 3-gram classifiers, showing improvements of 3\% and 9\%, respectively.
This result highlights the effectiveness of our approach.

\begin{figure*}[t]
    \centering
    \begin{subfigure}{0.4\textwidth}
        \begin{lstlisting}[language=python, breaklines=true]
n, m = map(int, input().split())
mi = 2
ma = n + m
ans = [1 for i in range(ma + 1)]

for i in range(2, int(ma**0.5) + 1):
    for j in range(i + i, ma + 1, i):
        ans[j] = 0
ans[0] = 0
ans[1] = 0
print(ans.count(1))



\end{lstlisting}
        \caption{Human code}
        \label{fig:correctly_predicted_human_solution}
    \end{subfigure}\hspace{3em}
    \begin{subfigure}{0.45\textwidth}
        \includegraphics[width=\linewidth]{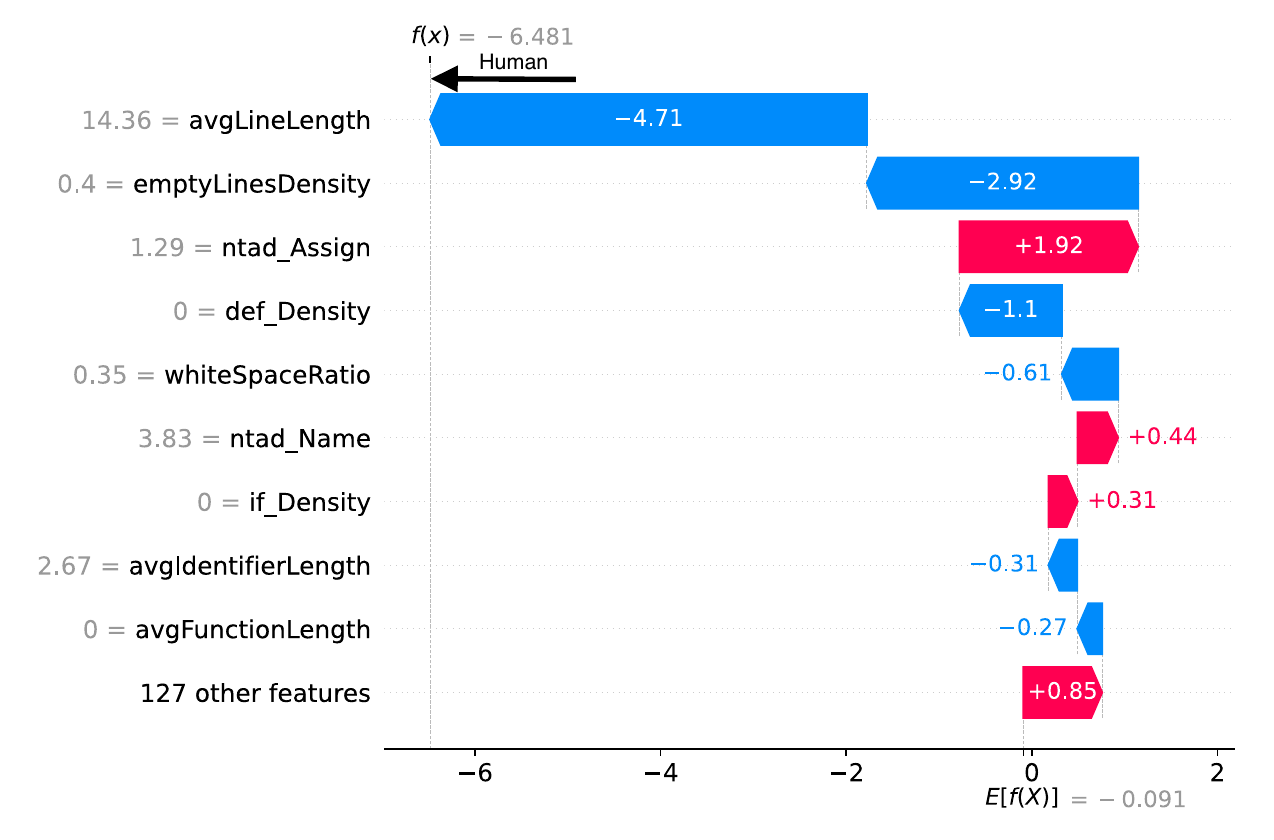}
        \caption{SHAP waterfall plot showing how features impact the model's decision for this code }
        \label{fig:wf_correctly_predicted_human_soln}
    \end{subfigure}
    \caption{Correctly predicted human code}
    \label{fig:correctly_predicted_human_solution_and_wfp}
\end{figure*}

\begin{figure*}[t]
    \centering
    \begin{subfigure}{0.4\textwidth}
        \begin{lstlisting}[language=python, breaklines=true]
def solve(n, m):
    return 1 if min(n, m) > 1 else 2

n, m = map(int, input().strip().split())
print(solve(n, m))\end{lstlisting}
        \caption{\aig code}
        \label{fig:correctly_predicted_ai_solution}
    \end{subfigure}\hspace{3em}
     \begin{subfigure}{0.45\textwidth}
        \includegraphics[width=\linewidth]{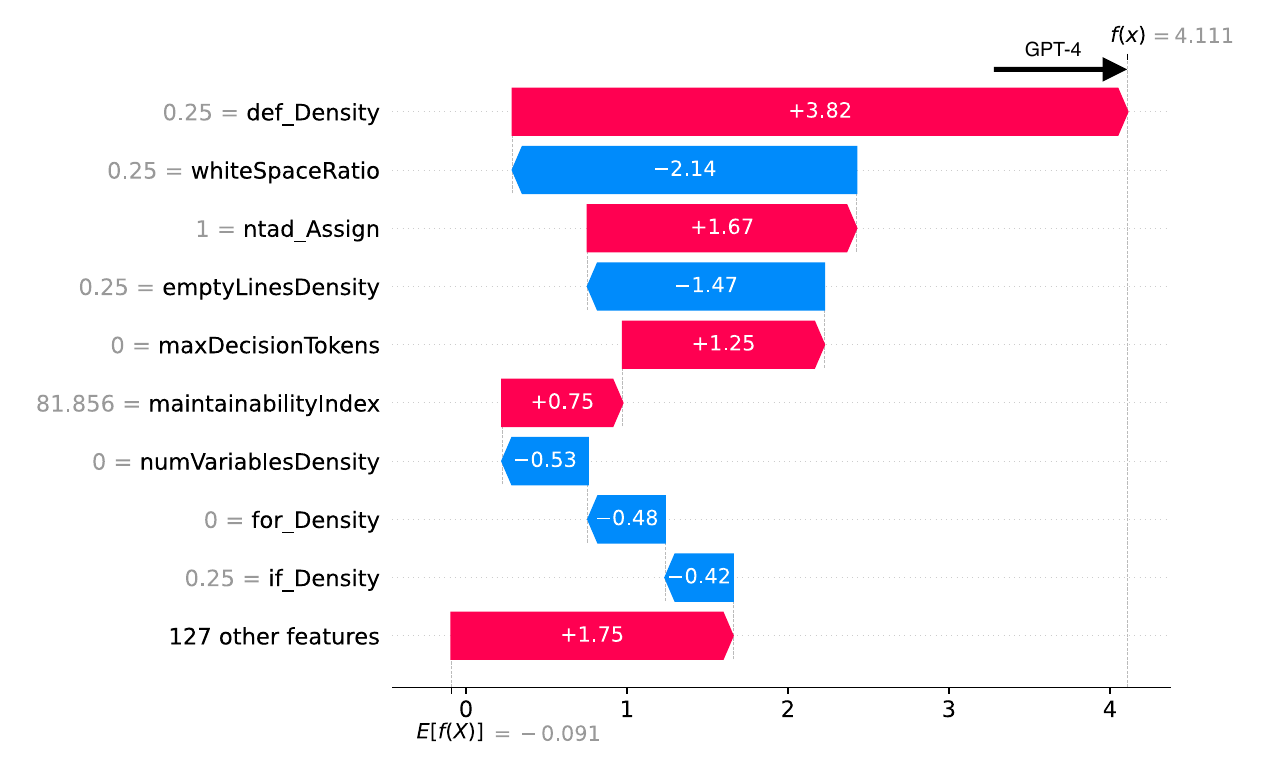}
        \caption{SHAP waterfall plot showing how features impact the model's decision for this code}
        \label{fig:wf_correctly_predicted_ai_soln}
    \end{subfigure}
    \caption{Correctly predicted \aig code.}
    \label{fig:correctly_predicted_ai_solution_and_wfp}
\end{figure*}

\begin{figure*}
    \centering
    \begin{subfigure}{0.485\textwidth}
       \begin{lstlisting}[language=python, breaklines=true]
def getCount(h,m,i) :
    h=int(h)
    m=int(m)
    h1=0
    lst=[11,22,33,44,55,66,77,88,99]
    while(h1<h) :
        for m1 in range(0,m) :
            if h1<10 :
                if(m1<10 and h1==m1) : count[i]+=1
                if (m1 in lst and m1%10==h1) :  count[i]+=1
            else :
                if(m1 in lst and h1==m1) : count[i]+=1
                if(h1 in lst and h1%10==m1) : count[i]+=1
        h1+=1
t=int(input())
count=[0]*t
for i in range(0,t) :
    h,m=input().split()
    getCount(h,m,i)

for i in count : print(i)\end{lstlisting}
    \caption{Human code}
    \label{fig:wrongly_predicted_human_code}
    \end{subfigure}\hspace{3em}
     \begin{subfigure}{0.45\textwidth}
        \includegraphics[width=\linewidth]{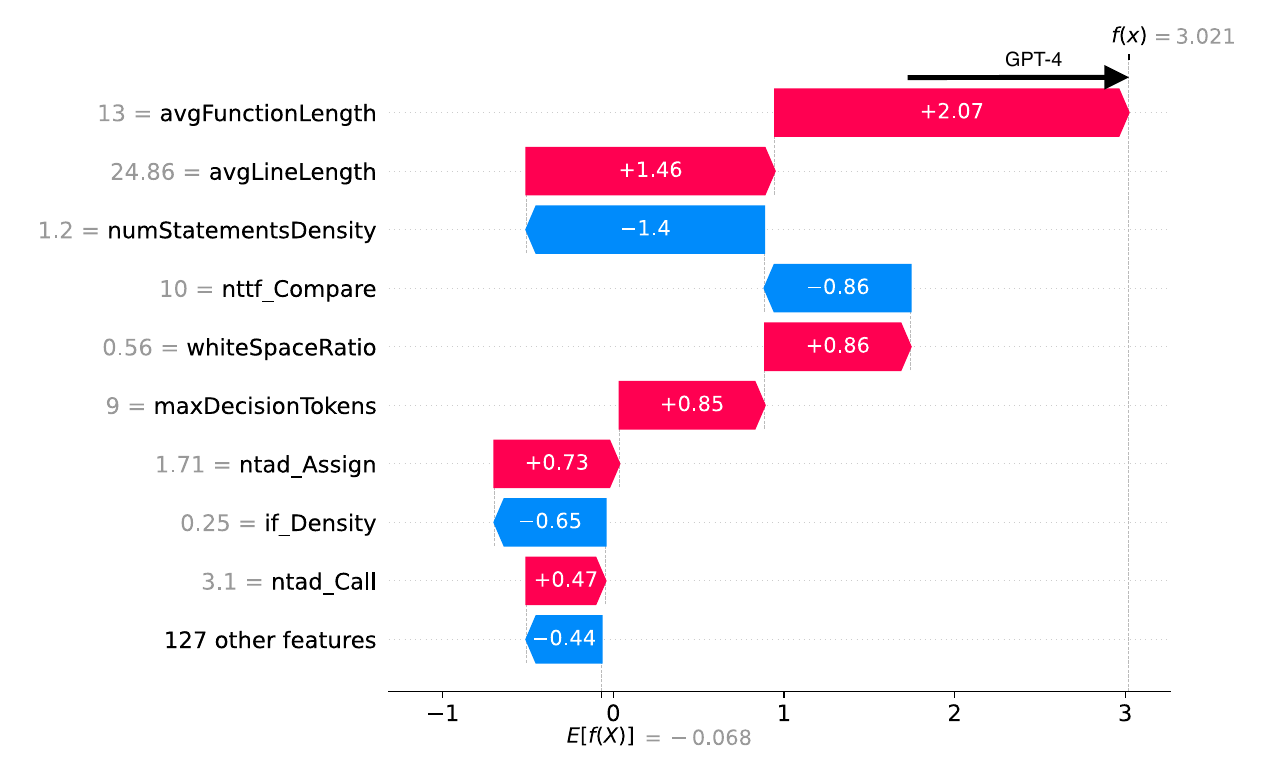}
        \caption{SHAP waterfall plot showing how  features impact the model's decision for this code}
        \label{fig:wf_wrongly_predicted_human_soln}
    \end{subfigure}
    \caption{Human code predicted incorrectly as AI-generated code}
    \label{fig:wf_wrongly_predicted_human_soln_and_wfp}
\end{figure*}

\section{Discussion}

Based on our findings, we make observations on our classifier, examining why it correctly and incorrectly predicts solutions in specific {instances. Understanding the reason behind its performance on a specific instance is crucial for humans to make the final decision whether it is human-authored or \aig. 

\subsection{Correctly Predicted Solution}

Figures \ref{fig:correctly_predicted_human_solution_and_wfp} and \ref{fig:correctly_predicted_ai_solution_and_wfp} present examples where our model correctly classifies (i.e., correctly distinguishes between human-authored and \aig code). Examining the important features in Figure \ref{fig:summary_plot_oa} and \ref{fig:summary_plot_non_gameable}, we can gain insights into the model's overall decision-making process by investigating the waterfall plot for each example. The waterfall plot presents the model's expected value, with each row representing how each feature contributes positively (red) towards \aig code or negatively (blue) towards human-authored code.

Figure \ref{fig:correctly_predicted_human_solution_and_wfp}, shows a correctly predicted 
human-authored code and its corresponding SHAP waterfall plot. This plot provides a local explanation of the classifier, visually depicting the key features that influence individual predictions, starting from the base value. At the bottom of the plot in Figure \ref{fig:wf_correctly_predicted_human_soln}, the model's prediction begins at this base value. It progresses towards the \aig class, influenced by the cumulative effect of 127 other features whose contributions are relatively minor and thus aggregated in the plot. These are ranked in ascending order of SHAP values. The \textit{ntad\_Assign} feature, representing the AST node for assignment operations, has the most impact towards the \ai class, with its value of 1.29. The value of this feature, in addition to the value of other features with red arrows in the plot, is indicative of the \aig class. However, the most impactful feature overall is the \textit{avgLineLength}, its value of 14.36 in addition to the value of other features with blue arrows in the plot influences the model's decision towards its final prediction, the human class. This shows that code having an \textit{avgLineLength} (the average length of characters in each line) of 14.36, \textit{emptyLinesDensity}, (i.e. the ratio of empty lines to sloc) of 0.4, \textit{def\_Density} of 0, \textit{whiteSpaceRatio} (i.e. the ratio of whitespace to non-whitespace characters) of 0.35, \textit{avgIdentifierLength} (i.e. the average length of identifier names) of 2.67 and no function is likely human-authored, not taking into consideration the other features.

Figure \ref{fig:correctly_predicted_ai_solution_and_wfp}  shows a correctly predicted \aig code and its corresponding SHAP waterfall plot. In Figure \ref{fig:wf_correctly_predicted_ai_soln}, the most important feature with a value of 0.25 is the \textit{def\textunderscore Density}. The value of this feature alongside the value of other features with red arrows influences the model's decision towards the final prediction, the \ai class. This suggests that code having \textit{def\textunderscore Density} of 0.25, \textit{ntad\textunderscore Assign} of 1, \textit{maxDecisionTokens} of 0, \textit{maintainabilityIndex} of 81.856 is likely \aig, not taking into consideration other features. MaintainabilityIndex is a feature that isn't observed by looking at the code, this is an interesting find on how a complexity metric could potentially influence the classifier's decision. 

These findings underscore the model's ability to correctly predict solutions based on distinctive features, shedding light on the significance of specific code characteristics in the classification process.

\subsection{Incorrectly Predicted Solution}

Figure \ref{fig:wf_wrongly_predicted_human_soln_and_wfp} presents an incorrectly predicted \aig code and its corresponding SHAP waterfall plot. This case is concerning, as the model incorrectly guesses the authorship of the code as \aig. 
While a false negative (predicting \aig code as human-authored) is bad, we especially want to avoid false positives. This is because we do not want to unjustly accuse someone of presenting \aig code when they have shared their written code. However, all classification techniques are bound to have some false positives. We hope that our choice of an explainable model helps educators look at the SHAP waterfall plot and understand the reasons before making a final decision as an educator. 

By examining important features from SHAP in Figure \ref{fig:summary_plot_oa} and \ref{fig:summary_plot_non_gameable}, we aim to gain insights into why the model made this incorrect prediction. 
In Figure \ref{fig:wf_wrongly_predicted_human_soln}, the two top features that influence the model's prediction towards the \aig class are related to the length of the code (i.e., avgFunctionLength, avgLineLength) and have values of 13 and 24.86, respectively.
In the case of both \textit{avgLineLength} and \textit{avgFunctionLength}, their values are high compared to the feature values of other observations within the dataset, and features with higher \textit{avgLineLength} and \textit{avgFunctionLength} tend to drive the model's prediction towards the \aig class. 

These observations explain why the model predicted this code as \aig, highlighting the challenges in accurately distinguishing certain code characteristics and the potential consequences of false positives.

\section{Threat to Validity}

We break down the threats into two parts, external and construct. 

\subsection{External Threat to Validity} 
These threats relate to the ability to generalize based on our results. 
In this study, we conducted an empirical investigation on the competitive programming platform CodeChef for human-authored code and utilized \ai for AI-generated code. 
Coding practices may vary on other platforms. However, in this study, to encompass different styles, we selected the most popular 100 problems for each difficulty level (i.e., from beginner level to 7* level) and extracted those that include Python solutions. This approach should enable us to generalize coding styles from beginners to experts. Our classifier performed well on easy, medium, and hard questions and we believe that due to the classifier's performance on easy and medium questions, it can perform fairly well on introductory programming courses. This is because problems in programming courses are likely to be of easy to medium difficulty. However, human solutions provided for easy to medium questions may not accurately represent the expertise level of all users. This means the solutions, while insightful, might not fully reflect the skill level of specific groups, such as students new to programming in introductory programming courses. Nonetheless, they do offer a general insight into human coding styles.
Another concern is since we only generate code using \ai, the generated code may not be representative of code by other AI assistants. However, compared to other tools, \ai is the most popular AI assistant and should represent real-world usage.
In future work, we aim to expand to other programming languages, and coding platforms and adapt to a broader range of AI models. 

\subsection{Construct Validity}
These threats relate to the degree to which our measurements are captured. Regarding the correctness of AI-generated code, we extract public test cases from CodeChef, which do not include private tests for the problems. We decided not to pursue direct submissions of AI-generated code to CodeChef, as this would violate ethical guidelines to submit AI-generated code as a human solution.

\vspace{-0.01cm}
\section{Conclusion}
The advent of AI assistants has introduced a new form of academic dishonesty, where students submit AI-generated code as their work. In this study, we investigated the impact of using code stylometry features to differentiate between human-authored and GPT-4 generated code, focusing on submissions from CodeChef and \aig solutions in Python. The findings demonstrate our approach's promise. Our classifier achieved an F1-score and AUC-ROC score of 0.91, highlighting its potential as a preliminary tool for identifying AI-generated code.
Additionally, we identified several key distinguishing features, with the average line length as the most important feature. By providing a means to identify \aig code, our study contributes to the ongoing discourse on the use and regulation of AI assistance in coding tasks.\\
\textbf{Future Work.} In the future, we could evaluate our hypothesis that our approach can perform as well on Python programs from other sources such as other competitive programming platforms and programming assignments from introductory courses. We could also extend our study to include other programming languages. Additionally, further research could evaluate the effectiveness of our approach in identifying AI-generated code that has been intentionally modified post-generation or through prompt engineering. Such investigation would provide valuable insights into the robustness and shortcomings of our detection methods when AI-generated code is deliberately altered to evade identification.
% \newpage
% \section{Acknowledgments}

%%
%% The next two lines define the bibliography style to be used, and
%% the bibliography file.
\bibliographystyle{ACM-Reference-Format}
\bibliography{msr_2024_ref}

\end{document}